\documentclass[12pt]{amsart}

\usepackage{graphicx}
\usepackage{amsmath}
\usepackage[pdftex]{hyperref}
\usepackage{graphics}
\usepackage{enumitem}
\usepackage[usenames,dvipsnames]{color}

\newtheorem{thm}{Theorem}
\newtheorem{lemma}{Lemma}
\newtheorem{prop}[thm]{Proposition}

\theoremstyle{definition}

\newtheorem{example}{Example}
\newtheorem{remark}{Remark}

\newcommand{\ignore}[1]{}

\begin{document}

\title{Models for transcript quantification from RNA-Seq}
\author{Lior Pachter}
\address{Departments of Mathematics and Molecular \& Cell Biology, UC Berkeley}
\email{\{lpachter\}@math.berkeley.edu}
\dedicatory{\today}

\begin{abstract}
RNA-Seq is rapidly becoming the standard technology for transcriptome
analysis. Fundamental to many of the applications of RNA-Seq is the
quantification problem, which is the
accurate measurement of relative transcript abundances from the
sequenced reads. We focus on this problem, and review many recently
published models that are used to estimate the relative
abundances. In addition to describing the models and the different
approaches to inference, we also explain how
methods are related to each other. A key result is that we show how inference with many of
the models results in identical estimates of relative abundances, even though model formulations
can be very different. In fact, we are able to show how a single
general model captures many of the elements of previously published
methods.  We also review the applications of RNA-Seq models to
differential analysis, and explain why accurate relative transcript
abundance estimates are crucial for downstream analyses.
\end{abstract}
\maketitle
\markboth{Lior Pachter}{Models for transcript quantification from RNA-Seq}


\section{Introduction}

The direct sequencing of transcripts, known as RNA-Seq \cite{Mortazavi2008}, has turned out
to have many applications beyond those of expression arrays. These
include genome annotation \cite{Denoeud2008}, comprehensive identification of fusions in
cancer \cite{Smith2010}, discovery of novel isoforms of genes \cite{Sultan2008}, and even
genome sequence assembly \cite{Mortazavi2010}. Moreover, RNA-Seq
resolution is at the level of individual isoforms of
genes \cite{Trapnell2010} and can be used to probe single cells \cite{Tang2009}. The wide
variety of applications of RNA-Seq \cite{Wang2009} require the solution of
many bioinformatics problems drawing on methods from
mathematics, computer science and statistics \cite{Pepke2009}.
The primary challenges are read mapping \cite{Trapnell2009b}, transcriptome assembly \cite{Haas2010},
quantification of relative transcript abundances from mapped fragment counts (the
topic of this review) and identification of
statistically significant changes in relative transcript
abundances when comparing different experiments \cite{Oshlack2010}. At this
point, it is therefore difficult, if not impossible, to survey the entire scope of work that
 constitutes RNA-Seq analysis in a single review. However
among the many challenges there is one that
is of singular importance: the accurate quantification of relative transcript
abundances. The hope is that RNA-Seq will be more accurate than
previous technologies, such as microarrays or even qRT-PCR, for inferring relative
transcript abundances \cite{Marioni2008}, and ultimately the success of RNA-Seq hinges on its
ability to deliver accurate abundance estimates.

We therefore focus on the problem of relative transcript abundance quantification in this
review and begin with a remark about what it means to quantify
abundances with RNA-Seq data.
\begin{remark}[Meaning of quantification for RNA-Seq] Since RNA-Seq
  consists of sequencing RNA (rather than protein), the technology
  does not measure what is technically {\em gene
    expression}. The term ``expression'' refers to the process by
  which functional products are generated from genes, and although in
  some cases a gene may consist of a non-coding RNA, the abundance of
  protein coding genes is mediated by translation \cite{Ingolia2009}.
 Even in the case where polyA selection is performed
  to enrich for mRNA that will be translated, what is measured in
  RNA-Seq are  the relative amounts of RNA {\em transcripts}. 

It is also important to note that RNA-Seq does
  not allow for the measurement of absolute transcript
  abundances. Because molecules are sampled proportionately, it can
  only be used to infer {\em relative transcript abundances}. 
\end{remark}

We are able to describe currently used methods in terms of a single common framework that explains how they
are all related. Our main result, outlined in Section 2, is the observation that the models
underlying existing methods can be viewed as special cases (or close
approximations) of a single model described in
\cite{Roberts2011}. In Section 3 we describe the simplest
class of models known as ``count based models''. In Section 4 we
focus on models for the estimation of individual relative transcript
abundances, and introduce a recurring theme which is the equivalence
of 
multinomial and Poisson log-linear models with respect to maximum
likelihood computations. We continue in Section 5 by describing a
general model  for RNA-Seq analysis that specializes to many previously published models.
Next, in Section 6, we discuss inference and parameter estimation in RNA-Seq
 models and in Section 7 we examine the applications of such estimates
 to the comparison of relative transcript abundances from two or more
 experiments. We conclude in Section 8 with a discussion and comments
 on speculations about future developments.

We have strived to minimize and simplify notation wherever possible, yet
have had to resort to defining many variables and parameters. To
assist the reader, Appendix II contains a glossary of notation used. 

\section{The RNA-Seq model hierarchy}

Stochastic models of RNA-Seq experiments underlie all methods for
obtaining relative transcript abundance estimates. In some cases, the
underlying models are only implicitly described. For example, this is
the case in
``count-based'' methods, where the total number of reads mapping to a region
(normalized by length and total number of reads in the experiment) is
used as a proxy for abundance. As we show in the next section, this intuitive measure is based on an
underlying model that is multinomial, and the normalized counts can be understood as the maximum likelihood estimate of parameters
corresponding to relative transcript abundances based on the model.

\begin{figure}[!h]
\includegraphics[scale=0.65]{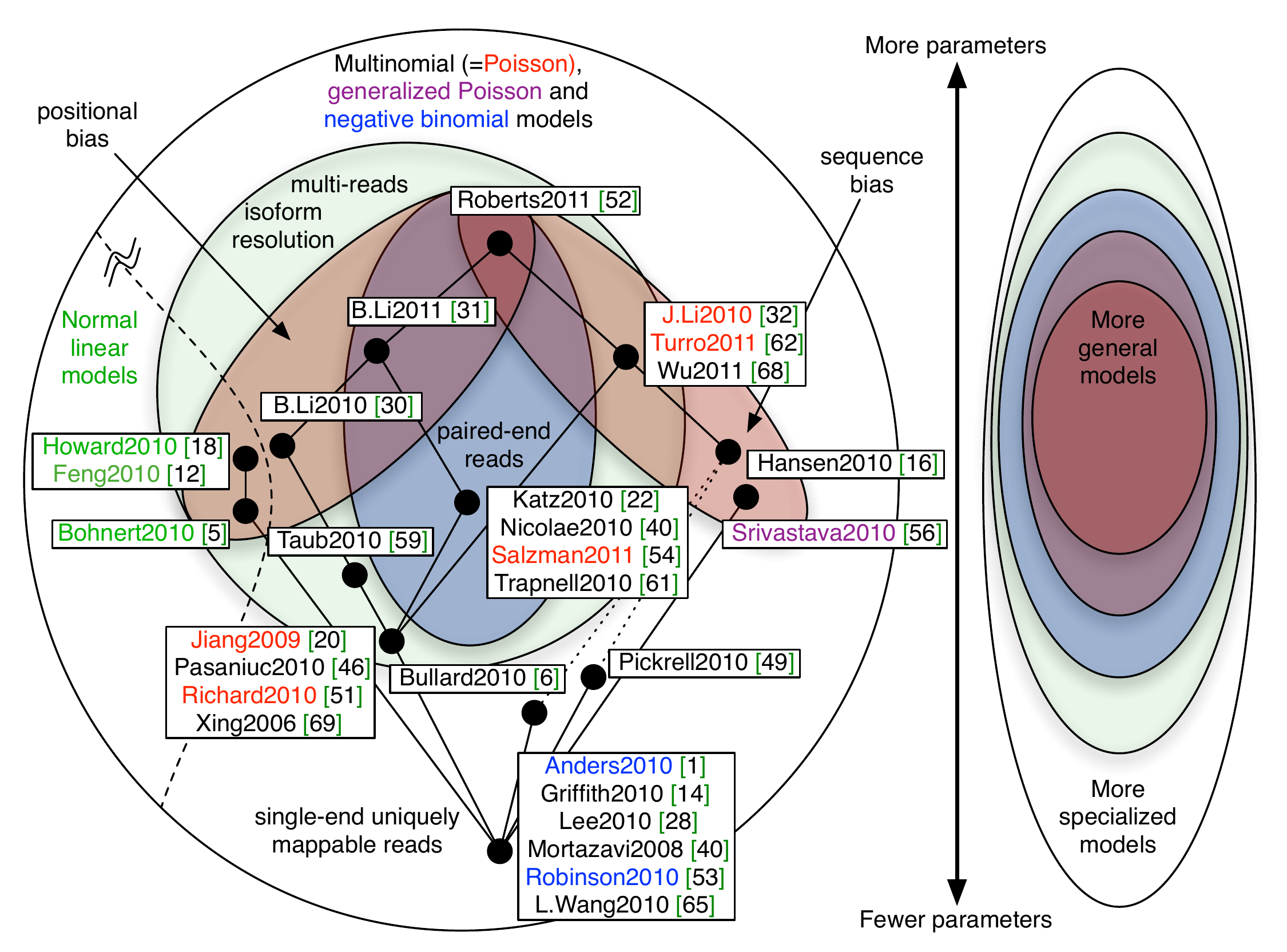}
\caption{Models for RNA-Seq. The figure shows a Venn diagram (and the
  partially ordered set it induces) representing relationships among models. More general models are
  nested inside simpler models. Models in the same box are
  mathematically equivalent and the boxes are organized so that more
   complex models (i.e., with more parameters) are positioned
  above simpler models with fewer parameters. Model types are color
  coded and the dashed line separates multinomial, Poisson and
  generalized Poisson models and negative binomial models from normal
  linear models. This is because the MLE obtained using the normal
  linear models in the multi-read case approximates the MLE obtained
  using the Poisson or multinomial models. Two models
  \cite{Bullard2010b,Pickrell2010} are connected with dashed lines to
  \cite{Hansen2010} because they include normalization steps that are
  related to sequence bias correction, but they are not strictly special cases
  of \cite{Hansen2010}. Some models, such as \cite{B.Li2010,Taub2010}, include
  terms for modeling errors in mapping that in principle can be adapted to model any
  biases/features (meaning that the models are in some sense the most
  general possible) but if applied in that way would be impractical (see
Remark \ref{rem:error}) which is why they are not at the top of the hierarchy.}
\label{fig:hierarchy}
\end{figure}

We have organized the published models for RNA-Seq analysis and
display the relationships among them in Figure \ref{fig:hierarchy}. Each node in the graph corresponds to a model for RNA-Seq, and
models are nested so that if there is a descending path from one node to another, then the latter model is a special case of the
former. In other words, the figure shows a partially ordered set
depicting relationships among models. Nodes are labeled by published
methods (first author+year+citation), and the shaded ellipses describe the
features modeled. The complexity of a model corresponds to the number
of ellipses it is contained in. Features modeled include:

\newsavebox{\mysquarez}
\newsavebox{\mysquarea}
\newsavebox{\mysquareb}
\newsavebox{\mysquarec}
\newsavebox{\mysquared}
\savebox{\mysquarez}{\textcolor{Black}{\rule{0.1in}{0.1in}}}
\savebox{\mysquarea}{\textcolor{LimeGreen}{\rule{0.1in}{0.1in}}}
\savebox{\mysquareb}{\textcolor{RoyalBlue}{\rule{0.1in}{0.1in}}}
\savebox{\mysquarec}{\textcolor{Thistle}{\rule{0.1in}{0.1in}}}
\savebox{\mysquared}{\textcolor{Thistle}{\rule{0.1in}{0.1in}}}

\begin{itemize}
\item count based models: the node at the bottom, contained only in the
``single-end uniquely mappable reads'' ellipse, refers to single end reads (i.e. not paired-end
reads) models in which all transcripts have a single
isoform and reads are uniquely mappable to transcripts. In the
simplest instance of such models there is no modeling of bias. 
\item  multi-reads (isoform resolution): these are models for individual relative transcript
  abundances in the case where reads may be sequenced from
  transcripts in genes with multiple isoforms. Equivalently, these are
  models for ``multi-reads'' which are reads that map to more than one
  transcript (not necessarily from the same gene). The first such model was proposed in \cite{Xing2006}
  and is equivalent to later formulations in \cite{Jiang2009,Pasaniuc2010,Richard2010}.
\item paired-end reads: these are models that include specific
  parameters for the length distribution of fragments. This is
  relevant when considering paired-end data in which the reads that
  form mate-pairs correspond to the ends of fragments that are being
  sequenced. Such models require estimation of fragment length
  distributions. The first paired-end model was published in
  \cite{Trapnell2010} and they subsequently appeared in \cite{Feng2010,Katz2010,Nicolae2010,Salzman2011}
\item positional bias: this refers to the non-uniformity of fragments
  along transcripts, and has been hypothesized to be the result of
  non-uniform fragmentation during library preparation. It was first modeled
  in \cite{B.Li2010,Bohnert2010,Howard2010} and later in \cite{Wu2011}
  but only for single-end reads. The model in \cite{B.Li2010}  was extended to a paired-end
  model in \cite{B.Li2011}. 
\item sequence bias: it has been empirically observed that sequences
  around the beginning and end of fragments are non-random leading to
  hypotheses that priming and fragmentation strategies bias
  fragments \cite{Hansen2010}. Models with parameters for sequence
  bias are \cite{J.Li2010,Turro2011}. In \cite{Roberts2011} both
  sequence bias and positional bias are modeled. It should be noted
  that sequence bias has been modeled indirectly, as in
  \cite{Pickrell2010} using GC content as a proxy (see
  \cite{Roberts2011} for more on the connection). 
\item In addition to the modeling of the specific features/effects
  discussed above, modeling of errors in reads is also discussed in
  some papers, e.g. \cite{Taub2010,B.Li2010}. It is not explicitly
  mentioned in Figure \ref{fig:hierarchy} because some models
  incorporate it implicitly in an {\it ad hoc} way by
  filtering during the mapping step (not discussed in this review). 
\end{itemize}

\begin{remark}[Mathematical equivalence of models]
Figure \ref{fig:hierarchy} is more than just a cartoon organizing the
models. Models from papers that appear in the same box are {\em
  mathematically identical} in terms of the quantification results
they will produce (although not necessarily when used for differential analysis, see
Section 7). We expand on this remark in the following
sections. However, it is is important to note that although models may
be equivalent, programs implementing them may not be. Implementation
details are important and non-trivial and can result in programs with
drastically different performance results and usability. For example, careful
attention to data types and processing of reads in a manner tailored
to the RNA-Seq protocol and sequencing technology used can improve quantification results. \end{remark}

\section{Count based models}
In this section we consider models that assume that all reads are
single-end and that they map uniquely
to transcripts. Such models are also known as ``count based
models''. In the simplest (generative) version of such a model, the
transcriptome consists of a set of transcripts with different
abundances, and a read is produced by
choosing a site in a transcript for the beginning of the read uniformly at random from among all
of the positions in the transcriptome.

Formally, if $T$
is the set of transcripts (with lengths $l_t, t \in T$) we define
$\rho=\{\rho_t\}_{t \in T}$ to be the {\em relative abundances
of the transcripts}, so that $\sum_{t \in T} \rho_t = 1$. We denote by 
 $F$ the set of (single-end) reads and let
 $F_t \subseteq F$ be the set of reads mapping to
transcript $t$. Furthermore, we use assume that all the reads in $F$
have the same length $m$. Note that in transcript $t$, the number of
positions in which a read can start is $\tilde{l}_t=l_t-m+1$. The
adjusted length $\tilde{l}_t$ is called the {\em effective length} of $t$.

In the generative model, first a transcript is chosen from which to
select a read $f$ by
\begin{equation}
\mathbb{P}(f \in t) \quad = \quad \frac{\rho_t \tilde{l}_t}{\sum_{r \in T} \rho_r
  \tilde{l}_r}.
\end{equation}
Next, a position in that transcript is selected uniformly at random
from among the $l_t-m+1$ positions. Thus, the likelihood of observing the reads $F$ as
a function of the parameters $\rho$ is 
\begin{equation}
\mathcal{L}(\rho) \quad = \quad \prod_{t \in T} \prod_{f \in F_t} \left( \frac{\rho_t \tilde{l}_t}{\sum_{r \in T} \rho_r
  \tilde{l}_r} \cdot \frac{1}{\tilde{l}_t}\right) .
\end{equation}
If we denote by $X_t$ the number of reads mapping to transcript $t$,
i.e. $X_t = |F_t|$, then we can rewrite the likelihood function as
\begin{equation}
\mathcal{L}(\rho) \quad = \quad \prod_{t \in T} \left(  \frac{\rho_t \tilde{l}_t}{\sum_{r \in T} \rho_r
  \tilde{l}_r}  \cdot \frac{1}{\tilde{l}_t}\right)^{X_t}.
\label{eq:singleread}
\end{equation}
There is a convenient change of variables that reveals Equation
\ref{eq:singleread} to be that of a log-linear model. 
Let $\alpha=\{\alpha_t\}_{t \in T}$ denote the probabilities of
selecting a read from the different transcripts:
\begin{equation}
\label{eq:rhoalpha}
\alpha_t \quad := \quad \mathbb{P}(f \in t)\quad  =\quad  \frac{\rho_t
  \tilde{l}_t}{\sum_{r \in T} \rho_r
  \tilde{l}_r} .
\end{equation}
Note that 
\begin{equation}
\label{eq:alphacondition}
\sum_{t \in  T}\alpha_t=1 \mbox{ and } \alpha_t \geq 0 \mbox{ for all } t.
\end{equation}
Moreover,  by Lemma 14 in the Supplementary Material of
\cite{Trapnell2010}, for any probability distribution
$\{\alpha_t\}_{t \in T}$ (i.e., satisfying the conditions of Equation
\ref{eq:alphacondition}), we can find a probability distribution $\rho$ so that Equation
\ref{eq:rhoalpha} is satisfied by setting 
\begin{equation}
\label{eq:inverserho}
\rho_t \quad = \quad \frac{\frac{\alpha_t}{\tilde{l}_t}}{\sum_{r \in T}
    \frac{\alpha_r}{\tilde{l}_r} }.
\end{equation}
It follows that if Equation \ref{eq:singleread} is rewritten as
\begin{eqnarray}
\mathcal{L}(\alpha)  & = &  \prod_{t \in T} \left( \frac{\alpha_t}{\tilde{l}_t}
\right)^{X_t}\\
& \propto &  \prod_{t \in T}  \alpha_t^{X_t},
\end{eqnarray}
then the unique maximum likelihood solution for $\alpha$ can be
transformed to the (unique) maximum likelihood solution for 
$\rho$. That is, from
\begin{equation}
\hat{\alpha}_t  \quad =\quad  \frac{X_t}{N}
\end{equation}
where $N=\sum_{t \in T}X_t$ is the total number of mapped reads, we can
apply (\ref{eq:inverserho}) to obtain that 
\begin{eqnarray}
\hat{\rho}_t & = & \frac{\frac{\hat{\alpha}_t}{\tilde{l}_t}}{\sum_{r \in T} \frac{\hat{\alpha}_r}{\tilde{l}_r}}\\
& = & \frac{X_t}{N} \cdot \frac{1}{\tilde{l}_t} \cdot \left( \frac{1}{\sum_{r \in T}
  \frac{X_r}{N\tilde{l}_r}} \right)\\
& \propto & \frac{X_t}{\left( \frac{\tilde{l}_t}{10^3}\right) \cdot
  \left(\frac{N}{10^6}\right)}\cdot \left( \frac{1}{\sum_{r \in T}
  \frac{X_r}{N\tilde{l}_r}} \right)\\
& \propto & \frac{X_t}{\left( \frac{\tilde{l}_t}{10^3}\right) \cdot \left(\frac{N}{10^6}\right)}\label{eq:rpkm}.
\end{eqnarray}
Equation \ref{eq:rpkm} is the RPKM (reads per kilobase per millions of
reads mapped) formula with which to measure abundance from
\cite{Mortazavi2008} (with the exception that in \cite{Mortazavi2008}
$l_t$ is used instead of $\tilde{l}_t$). Note that RPKM can be viewed
as a {\em method} because
the term abbreviates the procedure of evaluating 
Equation \ref{eq:rpkm}. However the derivation above shows that RPKM
is better thought of as a {\em unit} with which to measure relative transcript
abundance because it is (up to a scalar factor) the maximum likelihood
estimate for the $\rho$. Moreover, the statistical derivation of
relative abundance in RPKM units reveals that effective length
should be used instead of length, and it is evident that abundance
estimates reported in RPKM units are not absolute, but rather relative.

\begin{remark}[Normalizing the total number of reads] 
The number $N$ used has been defined to be the number of {\em mapped}
reads however we note that one can replace it by the number of {\em
  sequenced} reads if an extra faux ``noise'' transcript is included in the
analysis as the source of the unmappable reads \cite{B.Li2010}. We use the term
``noise'' because unmappable reads may represent sequencing mistakes that
result in meaningless data. Also, additional normalization steps may be
applied to improve the robustness of relative transcript abundance
estimates because extensive transcription of even a single gene
can drastically affect RPKM values. To address this, quantile
normalization was proposed in \cite{Bullard2010} and is
implemented in a number of software packages for RNA-Seq analysis, e.g.\cite{Anders2010,Langmead2010,Trapnell2010}.
\end{remark}

The model described above is suitable for single isoform genes and is
therefore appropriate in organisms such as bacteria where there is no
splicing. However in higher Eukaryotes with extensive alternative
splicing, alternative promoters, and possibly multiple polyadenylation
sites, there may be ambiguity in the assignment of reads to
transcripts and more complex models are necessary. Nevertheless,
Equation \ref{eq:singleread} has been used for inference where reads
map to multiple isoforms of a gene 
by using one of two different approaches that we refer to as {\em
  projective normalization} and {\em restriction to uniquely mappable reads}. 

Projective normalization is an approach to estimating the relative
abundance of genes consisting of multiple transcripts (corresponding to
different isoforms) directly from the total number of reads mapping to the gene locus. The approach
is based on 
Equation \ref{eq:rpkm} but with $X_t$ replaced by the total number of
reads mapping to the gene, and $l_t$ replaced by the total length of all
the transcripts comprising the gene after projection into genomic
coordinates (i.e., the union of all transcribed bases as represented
in genomic coordinates). This approach has been used in a number of methods,
including \cite{Griffith2010,Kasowski2010,Pickrell2010}.
In \cite[Proposition 3, Supplementary Material]{Trapnell2010} it is proved that projective normalization
always underestimates relative gene abundances, and in \cite{Wang2010}
empirical evidence is provided demonstrating that estimation of
individual relative transcript abundances (by maximum likelihood) improves
the accuracy of relative transcript abundance estimates. That is, the type of model
presented in the next section improves on projective normalization.

Restriction to uniquely mappable reads consists of only utilizing reads that
map uniquely to features of interest. Such an approach can be used for
abundance estimation in multiple isoform genes by identifying unique
features of transcripts (e.g. splice junctions unique to an isoform)
and applying Equation \ref{eq:rpkm} where $X_t$ is replaced by the
read counts for that feature and $l_t$ is suitably adjusted for the
length of the feature. Restriction to uniquely mappable reads has the major
problem that valuable data may be omitted from analysis because
it cannot be mapped to a unique transcript feature. For example, in many transcripts, the
unique feature may consist of a single junction. The approach of
\cite{Lee2010} consists of the restriction to uniquely mappable reads via an
adjustment of the length $l_t$ in Equation \ref{eq:rpkm} whereas in
\cite{Morin2008} the correction is performed through adjustments to coverage.

\begin{remark}[Species abundance estimation in metagenomics is related to
  transcriptome analysis]
The single end read model discussed above is relevant to abundance
estimation in metagenomics, where the problem of estimating relative
abundances of genomes in a community is related to relative transcript
abundance estimation of single isoform genes. For example, a formula
equivalent to Equation \ref{eq:rpkm} appears in
\cite{Angly2009}. Another closely related problem is that of inferring
haplotype frequencies from pooled samples \cite{Long2011}. These
analogies between metagenomics, haplotype inference and transcriptomics can be extended
further. For example,  {\it de novo} transcriptome assembly
requires the assembly of related sequences as in metagenome
assembly. One key difference between metagenome and transcriptome
assembly is that sequences from related species are more divergent
than sequences of isoforms from a single gene.
\end{remark}
\section{Models for multi-reads: estimating isoform abundances}

As discussed in the previous section, RNA-Seq allows for the
estimation of the relative abundance of  transcripts, however, this requires an
extension of Equation \ref{eq:singleread} to allow for ambiguously
mapped reads. In this section we show that the model of
\cite{Xing2006} is equivalent (with the exception of a length factor) to the model of \cite{Jiang2009}, and
show that the basis of the equivalence applies to many other
models. 

We begin by considering the likelihood function in
\cite{Xing2006} because that paper is, to our knowledge, the first
publication of the inference model needed for transcript level
relative abundance estimation from RNA-Seq. In \cite{Xing2006} inference
of relative abundances using expressed sequence tags (ESTs) is
discussed, but in terms of the model there is no
difference between ESTs and RNA-Seq with the exception of assumptions
about whether EST counts scale with the length of transcripts, an issue
we comment on below. The
likelihood function that is derived describes the probability of obtaining $N$
reads from a transcriptome with $K$ transcripts. Using the notation of \cite{Xing2006}, we let
${\bf Y}=\{y_{i,k}\}_{i=1,k=1}^{N,K}$ be the
matrix defined by $y_{i,k}=1$ if read $i$ aligns to
transcript $k$ and $0$ otherwise. This matrix is called the {\em
  compatibility matrix} (for an example see Equation \ref{eq:compatibility}). The likelihood is then shown to be
\begin{equation}
\label{eq:Xing}
\mathcal{L}(\alpha)\quad =\quad \prod_{i=1}^N  \left( \sum_{k=1}^K y_{i,k}
  \frac{\alpha_k}{\tilde{l}_k} \right).
\end{equation}
Here $\alpha=(\alpha_1,\ldots,\alpha_K)$ is defined as in the previous section (in
\cite{Xing2006} $\alpha_k$ is denoted by $p_k$) and $k
\in\{1,\ldots,K\}$ indexes the transcripts of the gene. We note that the
denominator $\tilde{l}_k$ (length of isoform $k$) is missing in
\cite{Xing2006}, which can be interpreted as an assumption that the
frequency of ESTs from a transcript is directly proportional to its
abundance (independent of the length of the transcript). This
assumption makes sense based on the oligo(DT) priming strategy for
ESTs, however some papers have suggested that truncated cDNAs
contribute substantially to ESTs due to internal priming
\cite{Nam2002}. If this is the case then the addition of the length in
the denominator is warranted; in the next section we derive a more general model of
which Equation \ref{eq:Xing} is a special case, and in that derivation
the reason for the denominator will become apparent.
\begin{remark}[Length normalization]
The (deliberate) omission or missed inclusion of the denominator
$\tilde{l}_k$ in \cite{Xing2006} probably did not matter much in
practice because the denominator is not needed if the lengths of all
isoforms are equal. In that case, the likelihood function is changed by a
scalar factor and therefore the maximum likelihood estimates for the
incorrect and correct models will be the same. Since abundance
estimates were only evaluated qualitatively in \cite{Xing2006}, the
presence or absence of the denominator may not have changed results
much. On the other hand, the paper \cite{Pasaniuc2010} is specifically
about RNA-Seq and therefore the omission of the denominator is an
error. 
\end{remark}

\begin{figure}[!ht]
\includegraphics[scale=0.50]{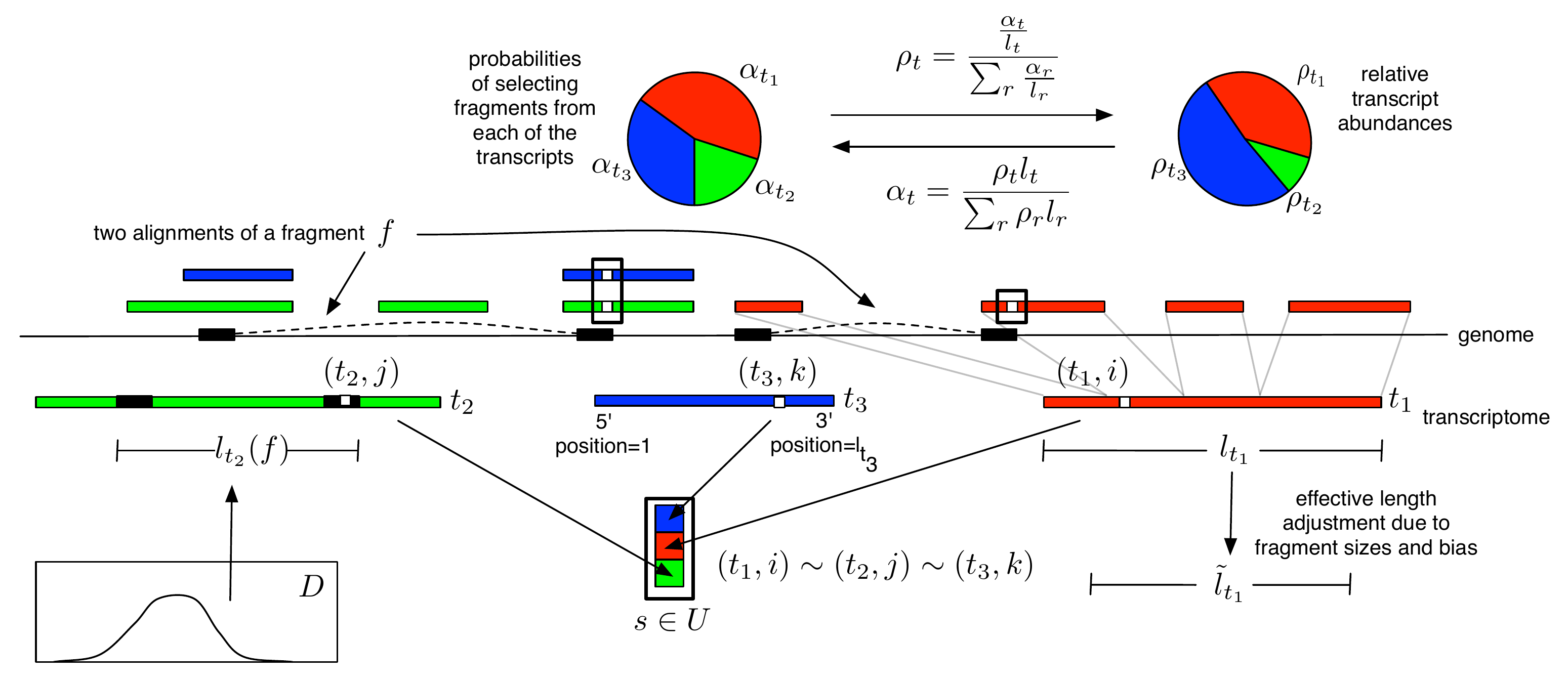}
\caption{Notation used in describing RNA-Seq models. In this example
  there are three transcripts ($t_1,t_2,t_3$) whose abundances are denoted
  by $\rho_{t_1},\rho_{t_2},\rho_{t_3}$. Fragment lengths are distributed
  according to $D$, and the length of a transcript $t$ is denoted by
  $l_{t}$. The effective length $\tilde{l}_t$ is an adjustment of the
  length taking into account bias and fragment length constraints.  Note that
  $l_{t_2}(f)$ is the length of the alignment of 
  fragment $f$ to transcript $t_2$, and may be different from $l_{t_1}(f)$. In the example, transcripts $t_2$ and $t_3$
  overlap in genomic coordinates. Three positions
  $(t_1,i),(t_2,j),(t_3,k)$ have been selected, one from each transcript, that lie in the same equivalence class
  (i.e. cannot be distinguished in mapping). This is indicated by
  $(t_1,i) \sim (t_2,j) \sim (t_3,k)$. Note that $(t_2,j) \sim
  (t_3,k)$ because they overlap in genomic coordinates whereas in the
  case of $(t_1,i)$ the transcript is part of a different gene.
  A fragment $f$ is shown mapping to
  transcript $t_2$.}
\label{fig:notation}
\end{figure}

To describe the model of \cite{Jiang2009}, we first introduce notation needed for multi-reads, which are
reads that may map to multiple transcripts. 
Note that there are two
reasons why reads may map to multiple transcripts: first, parts of
transcripts may overlap in genomic coordinates (in the case of
multi-isoform genes) and secondly, gene families may result in
duplicated segments throughout the genome that lead to ambiguous read
mappings (see Figure \ref{fig:notation}). In this section we restrict
ourselves to the ambiguous mappings resulting from multiple isoforms
of a single gene (to be consistent with the likelihood formulation of \cite{Xing2006}), but we introduce notation for the general case as it
will be useful later.

 Let $\mathcal{T}=\{(t,i):t \in T, i \in \{1,\ldots,l_t\}\}$ be the
set of all transcript positions. We first define a symmetric relation on
the set of positions by the relation $(t,i) R (u,j)$ if there
exists some fragment $f \in F$ so that the $3'$ end of $f$ aligns to
both $(t,i)$ and $(u,j)$. Note that if we allow for some mismatches in
alignments, the relation $R$ may not be an equivalence
relation. We will require the properties of equivalence relations in
specifying models so we replace $R$ by its equivalence closure (also called the
transitive closure) and denote the resulting equivalence relation by $\sim$.

The set of all equivalence
classes is the quotient of $\mathcal{T}$ by $\sim$ and we denote it by
$U=\mathcal{T}/\mathord\sim$. Given an equivalence class $s \in U$,
$F_s \subseteq F$  is the set of all fragments aligning to some
element in $s$ (note that this induces an equivalence relation on
fragments). For simplicity, we also assume that every fragment with a $3'$ alignment to some transcript
$t$ has only one alignment of its $5'$ end to a location in $t$.

In \cite{Jiang2009}, it is assumed that for a set of aligned fragments
$F$, for every $s \in U$ 
the number of reads starting at $s$ is Poisson distributed with rate parameter
\begin{equation}
\lambda_s \quad = \quad \sum_{k=1}^K c_{s,k} \frac{\kappa_k}{\tilde{l}_k},
\end{equation} 
where $\kappa_k$ is a rate parameter for transcript $k$. Here
${\bf C}=\{c_{s,k}\}_{k=1,s \in U}^K$ is a site-transcript
  compatibility matrix with $c_{s,k}=1$ if transcript $k$ appears in some element of $s$, and
$0$ otherwise. The parameters $\kappa_k$ are the expected number of
reads from each transcript $k$. If the observed number of reads starting at $s$ is
$X_s$, then the likelihood (Equation 2 in \cite{Jiang2009}) is given by
\begin{equation}
\label{eq:Jiang}
\mathcal{L}(\kappa)\quad  =\quad  \prod_{s \in U} \left( \frac{e^{-\lambda_s}\lambda_s^{X_s}}{X_s!}\right).
\end{equation}
At first glance Equations \ref{eq:Xing} and \ref{eq:Jiang} look very
different. The underlying models are distinct not just in name but in
the generative model for RNA-Seq that they imply. In the Poisson model
counts are Poisson distributed, which is only {\em approximately} the same
as the count distribution from the multinomial model. It is therefore
no surprise that the likelihood functions for the two different models have a different
form. Moreover, neither the parameters $\alpha$ nor $\rho$ appear in
Equation \ref{eq:Jiang}. However, we will see that Equation
\ref{eq:Jiang} can be written in terms of parameters equivalent to the
$\alpha$ and that the two likelihood functions are maximized at the same
value. This follows from an elementary and well-known equivalence between multinomial and
Poisson log-linear models \cite{Lang1996}. It is discussed in the
RNA-Seq case in \cite{Richard2010} but for completeness and clarity,
we review the connection between the models in detail below.

\begin{prop}
Maximum likelihood parameters obtained via Equations \ref{eq:Xing} and
\ref{eq:Jiang} lead to exactly the same relative transcript abundance
estimates. 
\end{prop}
{\bf Proof}:  Note that $\sum_{s \in U}X_s \mbox{ (from
  \cite{Jiang2009}) }= N \mbox{ (from \cite{Xing2006})}$ by definition. We will
also define parameters $\beta_k := \frac{\kappa_k}{N}$,
$Z=\sum_{k=1}^K\beta_k$ and $\gamma_k =
\frac{\beta_k}{Z}$. Note that $\sum_{k=1}^K \gamma_k = 1$ and
$\gamma_k \geq 0$, however since the $\kappa_k$ and $\beta_k$ are
unconstrained the factor $Z$ can, {\it a priori} be arbitrary. The
proof is based on the result that the maximum likelihood values for
$\kappa$ are obtained when $Z=1$. We derive this after proving a
simple combinatorial lemma that is useful inside the main argument. The lemma states
that each $\kappa_k$ is counted once when summing over the
equivalence classes of positions (and suitably normalizing by
length):
\begin{lemma}
\[\sum_{k=1}^K \kappa_k = \sum_{s \in U} \sum_{k=1}^K
c_{s,k}\frac{\kappa_k}{\tilde{l}_k}. \]
\end{lemma}
{\bf Proof}:
\begin{eqnarray}
{\sum_{s \in U} \sum_{k=1}^K c_{s,k}\frac{\kappa_k}{\tilde{l}_k}} & = & \sum_{k=1}^K \sum_{s \in U} c_{s,k}\frac{\kappa_k}{\tilde{l}_k}\\
& = & {\sum_{k=1}^K \tilde{l}_k \cdot \frac{\kappa_k}{\tilde{l}_k}}\\
& = & \sum_{k=1}^K \kappa_k. \qed
\end{eqnarray}
From this it follows that 
\begin{equation}
\sum_{s \in U} \sum_{k=1}^K c_{s,k}\frac{\kappa_k}{\tilde{l}_k} =N
\sum_{k=1}^K \beta_k
\end{equation}
which is used in the proof of the proposition: 
\begin{eqnarray}
\mathcal{L}(\kappa) & = & \prod_{s \in U}  \left(
  \frac{e^{-\lambda_s}\lambda_s^{X_s}}{X_s!} \right)\\
& = & e^{-\sum_{s \in s}\lambda_s} \prod_{s \in U}\left(
  \frac{\lambda_s^{X_s}}{X_s!} \right)\\
& \propto & e^{-\sum_{s \in U} \sum_{k=1}^K c_{s,k}\frac{\kappa_k}{\tilde{l}_k}}\prod_{s
  \in U}\left( \sum_{k=1}^Kc_{s,k} \frac{\kappa_k}{\tilde{l}_k}
\right)^{X_s}\\
& = &  e^{-N\sum_{k=1}^K \beta_k}\prod_{s \in U} \left( \sum_{k=1}^Kc_{s,k} N \frac{\beta_k}{\tilde{l}_k} \right)^{X_s} \\
& \propto &  \left( \frac{e^{-N\sum_{k=1}^K \beta_k}N^N}{N!} \right) \prod_{i=1}^N
\left( \sum_{k=1}^K y_{i,k}\frac{\beta_k}{\tilde{l}_k} \right) \\
& \propto & \left( \frac{e^{-NZ}(NZ)^N}{N!}\right) \prod_{i=1}^N  \left( \sum_{k=1}^K y_{i,k}\frac{\gamma_k}{\tilde{l}_k} \right).
\end{eqnarray}
A key observation is that the parameter $Z$ in the left parentheses can be
maximized independently of the $\gamma_k$ in the right
parentheses. This is because $Z$ is unconstrained and the constraints
on $\gamma$ do not involve $Z$ (the $\gamma$ only need to be non-negative and
must sum to 1). The maximum likelihood estimate for the
normalization constant $Z$ is $\hat{Z}=1$ and
therefore $\hat{\beta}_k=\hat{\gamma}_k$ and the maximization of Equation \ref{eq:Jiang} is equivalent to maximizing
\begin{equation}
\mathcal{L}(\beta) \quad = \quad \prod_{i=1}^N\left( \sum_{k=1}^K y_{i,k}\frac{\beta_k}{\tilde{l}_k} \right) 
\end{equation}
where $\sum_{k=1}^K \beta_k = 1$ and $\beta_k \geq 0$. The
interpretation of the $\beta$ parameters in the Poisson model
\cite{Jiang2009} is equivalent to the interpretation of the $\alpha$
parameters in \cite{Xing2006}, i.e. they are the
probabilities for choosing reads from transcripts. Therefore, both
model formulations are equivalent. \qed

\section{A general model for RNA-Seq}

In this section we present a general model for RNA-Seq that includes
as special cases the other models discussed in this review.
As before,  we assume that every fragment with a $3'$ alignment to some transcript
$t$, has only one alignment of its $5'$ end to a location in $t$. This
means that the notion of the length of a fragment relative to a
transcript is well-defined, and we denote this length by
$l_t(f)$. This assumption is slightly restrictive, but simplifies a
lot of the notation.

In a generative description of the model the $3'$ end of a fragment is
selected first. The probability of
selecting the $3'$ end from a specific site within a transcript
depends on the abundance of the site relative to others (as determined
by the relative abundance parameters $\rho$), on the local
sequence content, and also on the relative position of the site within
the transcript. Then a length for the fragment is selected according
to a distribution $D$ and again according to the local sequence content.

\begin{figure}[ht]
\includegraphics[scale=0.4]{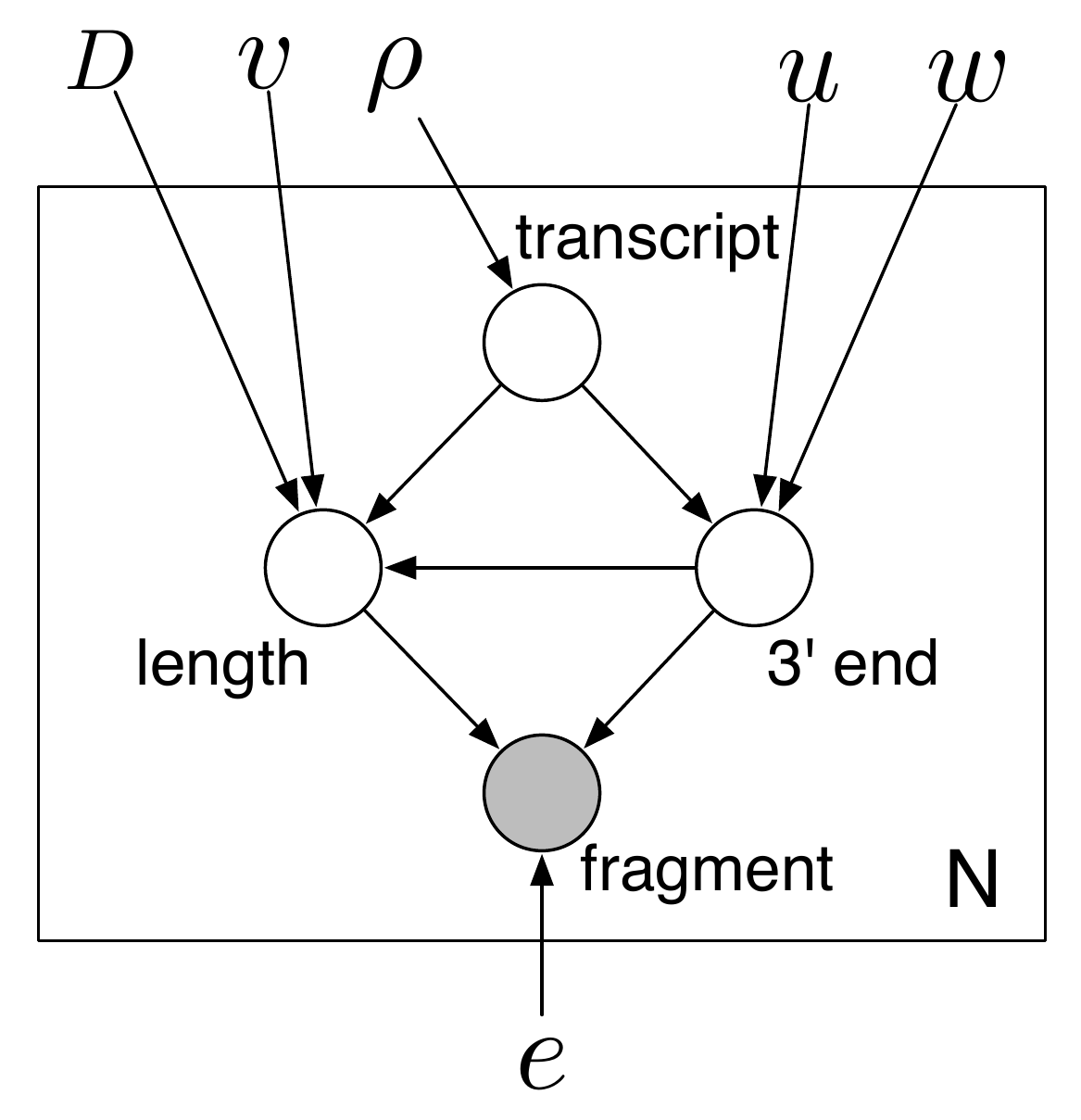}
\caption{A generative (graphical) model for RNA-Seq.}
\label{fig:model}
\end{figure}

We are
interested in estimating the $\rho_t$, but due to the other parameters
specifying a sequencing experiment we infer them indirectly. 
To specify the probability of a fragment with specific
$5'$ and $3'$ ends, we require the following parameters:
\begin{itemize}
\item The fragment length distribution denoted $D$ which is a distribution whose support
  is the positive integers. 
\item Site specific bias: parameters $u_{(t,i)}$ denoting the $3'$ site specific bias for position
$i$ in transcript $t$. These parameter are non-negative real numbers
(or ``weights''), so that no bias
corresponds to all $u_{(t,i)}=1$. Similarly, the $5'$ site specific bias
for position $i$ in transcript $t$ is denoted by $v_{(t,i)}$.
\item Positional bias: parameters $w_x$ where $x \in [0,1]$ that are
  non-negative real numbers.
\item Errors in reads: parameters $e_{t,f}$ denote the probability of
  observing the sequence in $f$ assuming that it was produced from
  transcript $t$. We assume (for simplicity) that every fragment could
  have been generated from at most one position in each
  transcript. The error probabilities $e_{t,f}$ can be
  estimated according to an error model for sequencing and based on
  the number and locations of mismatches/indels between $f$ and
  $t$. Note that $e_{t,f}$ can be viewed as a generalization of
  $y_{i,k}$ from the previous section where $i$ is replaced by $f$
  and $k$ by $t$ and the values form a probability distribution rather
  than being restricted to the set $\{0,1\}$.
\end{itemize}
Our model specifies that the probability of selecting a fragment $f$ with $3'$ end $(t,i)$ given
that $f$ originates from transcript $t$ is 
\begin{equation}
\mathbb{P}(f^{3'}  =  (t,i)|f \in t)\quad  =\quad   \frac{w_{\frac{i}{l_t}}\cdot u_{(t,i)} \cdot  \sum_{j=1}^{i-1}\frac{D(i-j)}{\sum_{k=1}^{i-1}D(i-k)} v_{(t,j)}}{\tilde{l}_t},
\end{equation}
where
\begin{equation}
\tilde{l}_t \quad = \quad \sum_{i \in t} \left( w_{\frac{i}{l_t}}\cdot u_{(t,i)}
\cdot \sum_{j=1}^{i-1}\frac{D(i-j)}{\sum_{k=1}^{i-1}D(i-k)} v_{(t,j)}\right).
\end{equation}
As in Section 3, $\tilde{l}_t$ is the effective length of transcript $t$.
The probability of selecting a fragment $f$ with $3'$ end $(t,i)$ is
now given by 
\begin{eqnarray}
\mu_{(t,i)} := \mathbb{P}(f^{3'} = (t,i)) &  =& \mathbb{P}(f \in t) \mathbb{P}(f^{3'} = (t,i)|f \in t) \\ 
& = & \alpha_t  \frac{w_{\frac{i}{l_t}}\cdot u_{(t,i)} \cdot  \sum_{j=1}^{i-1}\frac{D(i-j)}{\sum_{k=1}^{i-1}D(i-k)} v_{(t,j)}}{\tilde{l}_t},
\end{eqnarray}
and the conditional probability that a fragment $f$ with $3'$ end $(t,i)$ has length $l_t(f)$ is
\begin{equation}
\zeta^f_{(t,i)}\quad  :=\quad  \mathbb{P}(l(f) = l_t(f)|f^{3'}=(t,i))
\quad =\quad \frac{D(l_t(f)) v_{(t,i-l+1)}}{\sum_{j=1}^{i-1}D(i-j)v_{(t,j)} }.
\end{equation}
Therefore, the probability of generating a fragment $f$ of length $l_t(f)$ with $3'$ end
$(t,i)$ is 
\begin{eqnarray}
  \mathbb{P}(f^{3'} =  (t,i),l(f)=l_t(f)) & = & \mu_{(t,i)} \zeta^f_{(t,i)}\\
& = & \alpha_t \frac{ u_{(t,i)} \cdot v_{(t,i-l+1)} \cdot w_{\frac{i}{l_t}}
  \frac{D(l_t(f))}{ \sum_{k=1}^{i-1}D(i-k) } }{\tilde{l}_t}.
\end{eqnarray}
The generative model we have just described is summarized in Figure \ref{fig:model}.

We can now derive the likelihood for a set of fragments:
\begin{eqnarray}
\mathcal{L}(\alpha)& = &  \prod_{s \in U} \prod_{f \in
  F_s} \sum_{(t,i) \in s}\alpha_t e_{t,f} \mathbb{P}(f^{3'}=(t,i),l(f)=l_t(f)|f \in t)\\
& = & \prod_{s \in U} \prod_{f \in
  F_s} \sum_{(t,i) \in s}\mu_{(t,i)} \zeta^f_{(t,i)}\\
& = & \prod_{s \in U} \prod_{f \in
  F_s} \sum_{(t,i) \in s}
\alpha_te_{t,f}\frac{1}{\tilde{l}_t} \frac{D(l_t(f))}{\sum_{k=1}^{i-1}D(i-k) }
u_{(t,i)}  v_{(t,i-l_t(f)+1)} w_{\frac{i}{l_t}} .\label{eq:genlik}
\end{eqnarray}

This is a linear model (for fixed bias and error parameters) that is
concave. In the next section we discuss how maximum likelihood
estimates are obtained. Note that the inferred parameters $\hat{\alpha}$ can be
translated to maximum likelihood estimates for the relative
abundances $\hat{\rho}$ using Equation \ref{eq:inverserho}. Estimates are usually
reported in units of FPKM (fragments per kilobase per million
mapped fragments) \cite{Trapnell2010} and are a scalar multiple of
the $\hat{\rho}$ as in Equation \ref{eq:rpkm}. 

In the case when bias is not modeled,
i.e. $u_{(t,i)}=v_{(t,i)}=w_{(t,i)}=1$, then Equation \ref{eq:genlik}
reduces to Equation 9 in the Supplementary Methods of
\cite{Trapnell2010} (with the slight difference in model as explained
in Remark \ref{rem:length}). This is also the model in \cite{B.Li2010}
(with $\rho_t$ named $\theta_t$) and where the parameters $w_{(t,i)}$
are in the model but $u_{(t,i)}=v_{(t,i)}=1$. The paper \cite{B.Li2010}
describes a single read model and the paired-end case will appear in
\cite{B.Li2011}. This model is also the one used in
\cite{Pasaniuc2010} ($\alpha_t$ is named $P_t$, and formulated as an equivalent
Poisson model).

\begin{remark}[Directional assymetry in RNA-Seq]
\label{rem:length}
There is some confusion in the current literature about how to model
the generation of paired-end fragments in RNA-Seq experiments (e.g. in
\cite{Feng2010} three ``strategies'' are proposed). Some models
consider a process where the $5'$ end of a read is generated first
(usually uniformly at random) followed by the $3'$ end according to
the fragment length distribution (normalized appropriately if the $5'$
site is close to the $3'$ end of the transcript). In the model
described above, we have preferred to assume that the $3'$ site is
generated first, because that more closely mimics the actual
protocol. Alternatively, one can assume that first a length is chosen
according to the fragment length distribution, and then the $5'$ and
$3'$ sites are chosen \cite{Roberts2011}. Such a model may better
capture the fact that size selection follows reverse transcription and
double stranded cDNA generation. It is at present unclear which
formulation produces best estimates, but regardless of the choice the
effect on the likelihood function is to slightly alter the form of the
denominator. 
\end{remark}

\begin{remark}[Errors in reads] \label{rem:error} The error model we have included
(parameters $e_{t,f}$) is analogous to the formulation in
\cite{Taub2010} and allows for a general and position-specific
modeling of errors. However, it should be noted that there is a
connection between errors and read mapping that can skew results even
when errors are modeled. A read with a number of errors beyond the
threshold of the mapping program used is not mapped, and this missing
data problem is not addressed in our model. This issue can affect
expression estimates, especially in the case of allele specific
estimation \cite{Degner2009} (see also remark below). It is therefore
advisable to correct errors before mapping, using methods such as in
\cite{Yang2011}. In \cite{Taub2010} {\em all} possible mapping are
considered, i.e. the entire read-transcript compatibility matrix is
constructed, and while this is the most general model
  for RNA-Seq since in principle the ``error'' model can capture any
  feature,  it is impractical because the compatibility matrix is too large to work with explicitly in
  practical examples.
\end{remark}
 
\begin{remark}[Allele specific estimates] It may be desirable to
  infer relative transcript abundances for individual {\em haplotypes} and this
 problem is addressed in various methods papers
 \cite{Bullard2010b,Montgomery2010,Pasaniuc2010,Pickrell2010,Turro2011}. It is possible using the formalism we have described by simply doubling
  the number of transcripts (one for each haplotype) and utilizing the
  error parameters $e_{t,f}$ to obtain probabilities for each fragment
  originating from each of the haplotypes. Note that if the haplotypes
  are unknown then heterozygous sites can be inferred from the mapped
  fragments, but we do not consider that problem in this review as it
  is a mapping issue rather than a modeling problem.
\end{remark}

The model we have presented is multinomial, however as discussed
previously it can be formulated as an equivalent Poisson model. The
details, and a proof that the two formulations are equivalent, is
provided in Appendix I.

\section{Inference}

The single read single isoform model described in Section 2 is log-linear and
therefore admits a closed form solution. Models that allow for
ambiguous read mapping are no longer log-linear, but they do have nice
properties and, assuming that bias effects are known, are
concave \cite{Jiang2009,Trapnell2010}. This means that numerical
algorithms can be used to find the (unique) global maximum assuming
that the model is identifiable.

\begin{remark}[Identifiability]
Identifiability of RNA-Seq models is addressed in \cite{Lacroix2008,Hiller2009}
and appears even earlier in the computational biology literature in
related models used in other applications
(e.g.,\cite{Peer2004}). Identifiability is the statistical property of
``inference being possible''. Mathematically, that means that
different parameter values (relative transcript abundances) generate
different probability distributions on read counts. Testing for
identifiability in our setting is equivalent to
determining whether the compatibility matrix (Section 4) 
is full rank \cite{Hiller2009}.
The identifiability
problem is related to the transcript assembly problem, because with
certain assumptions assemblies can be guaranteed to generate
identifiable models for the aligned reads (this is the case with Cufflinks assemblies
\cite{Trapnell2010}). 
\end{remark}

Typically the Expectation-Maximization (EM) algorithm is used for
optimization because of its simplicity in
formulation and implementation. 

\begin{figure}[!ht]
\includegraphics[scale=0.62]{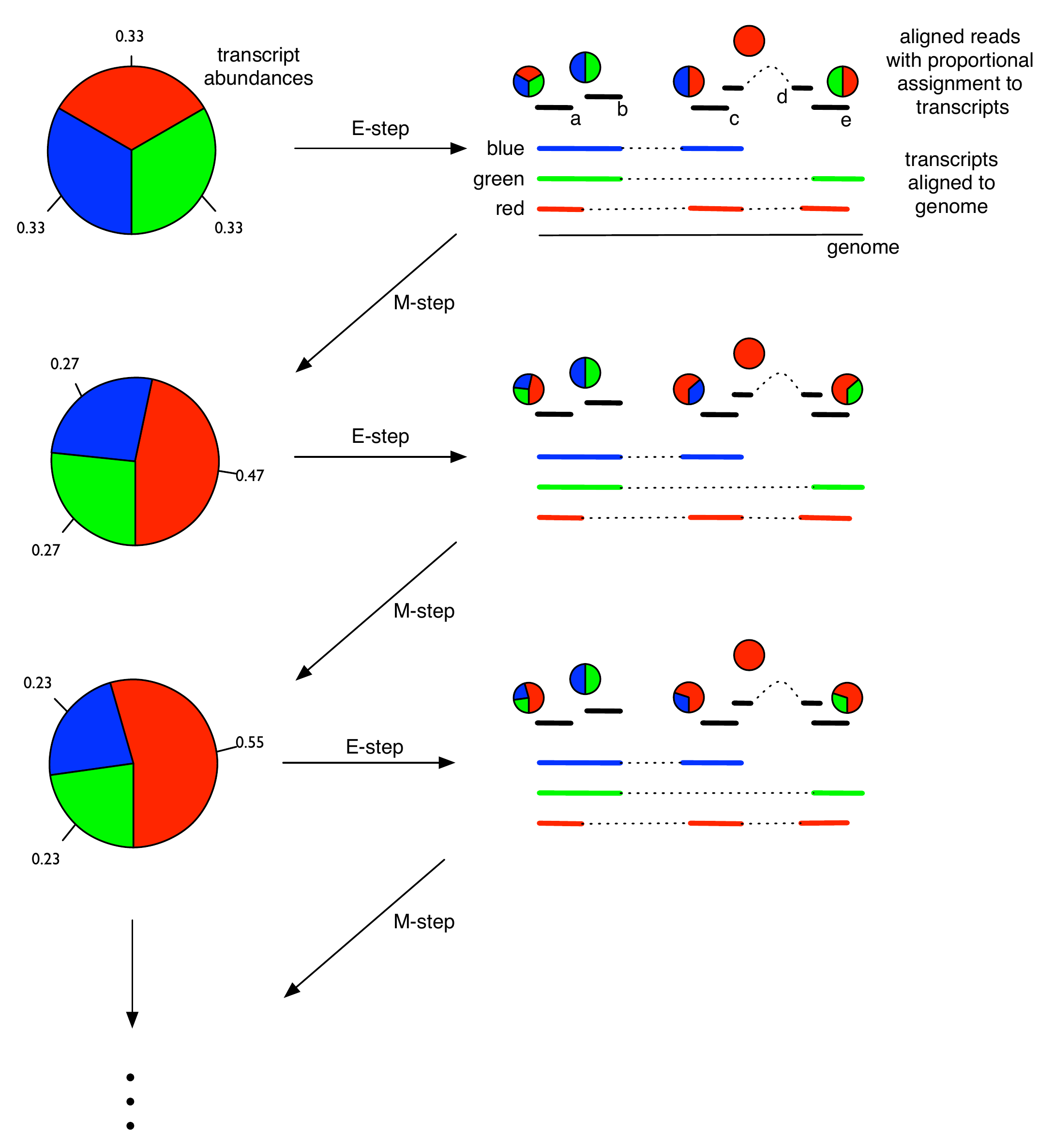}
\caption{Illustration of the EM algorithm. The gene has
  three isoforms ({\tt red, green, blue}) of the same length. There are five
reads (a,b,c,d,e) mapping to the gene. One maps to all three isoforms, one only to
red, and the other three to each of the three pairs of isoforms.
 Initially every isoform is assigned the same abundance
($\frac{1}{3},\frac{1}{3},\frac{1}{3}$). During the {\em expectation} (E)
step reads are proportionately assigned to transcripts according to
the isoform abundances. Next, during the {\em maximization} (M) step
isoform abundances are recalculated from the proportionately assigned
read counts. Thus, for example, the abundance of {\tt red} after the
first $M$ step is estimated by \newline $0.47=(0.33+0.5+1+0.5)/(2.33+1.33+1.33)$.}
\label{fig:em}
\end{figure}

\begin{example}
Consider a gene with three transcripts of equal length labeled  {\tt red},
{\tt green} and {\tt blue}, and with 5 single-end reads aligning to
the transcripts according to the configuration shown in Figure \ref{fig:em}. If the reads are labeled $a,b,c,d,e$ then the compatibility matrix (see Section 4) is
\begin{equation}
\label{eq:compatibility}
{\bf Y}\quad =\quad \bordermatrix{
          & a & b & c & d & e\cr
{\tt red}     & 1 & 0 & 1 & 1 & 1\cr 
{\tt green} & 1 & 1 & 0 & 0 & 1\cr
{\tt blue}   & 1 & 1 & 1 & 0 & 0\cr
}.
\end{equation}
If the transcripts have
abundances $\rho_{\mbox{\tt red}},\rho_{\mbox{\tt
    blue}},\rho_{\mbox{\tt green}}$
then the estimation of the $\rho$ (using the model in Section 3) is the mathematics problem of
maximizing the likelihood function
\begin{equation}
\mathcal{L}(\rho) \quad  =\quad 
(\rho_{\mbox{\tt red}}+\rho_{\mbox{\tt blue}})(\rho_{\mbox{\tt
    red}}+\rho_{\mbox{\tt green}})(\rho_{\mbox{\tt
    blue}}+\rho_{\mbox{\tt green}})\rho_{\mbox{\tt red}}
\end{equation}
subject to the constraint $\rho_{\mbox{\tt red}}+\rho_{\mbox{\tt
    blue}}+\rho_{\mbox{\tt green}}=1$. 
First, we note that the matrix $Y$ is full rank (rank=$3$) and therefore the model is identifiable.
\begin{prop}
The maximum
likelihood solution is given by 
\begin{equation}
\label{eq:exmle}
\hat{\rho}_{\mbox{\tt blue}}  \quad =\quad  \hat{\rho}_{\mbox{\tt green}}\quad =
\quad \frac{7}{16} -
\frac{\sqrt{17}}{16} \quad \approx \quad 0.18.
\end{equation}
\end{prop}
{\bf Proof}: For notational simplicity we let $x=\rho_{\mbox{\tt blue}}, y=\rho_{\mbox{\tt green}},z=\rho_{\mbox{\tt
    red}}$. The
goal is to maximize $f(x,y,z)=z(1-x)(1-y)(1-z)$ subject to the
constraint that $x+y+z=1$ and $x,y,z \geq 0$. Note that 
$f(x,y,z)=(x+y)(1-(x+y))(1+xy-(x+y))$. For any value $x+y$ at which $f$
is maximized, it follows that $xy$ must be maximized conditional on
the sum, and that occurs at $x=y$. Therefore, the problem is
equivalent to maximizing $2x(1-2x)(1+x^2-2x)$ where $0 \leq x
\leq \frac{1}{2}$. The maximum is at $\hat{x} = \frac{7}{16} - \frac{\sqrt{17}}{16}$.\qed

In the EM algorithm,
relative transcript abundances are first initialized (e.g. to the uniform
distribution). In the expectation step every read (or fragment in the
case of paired-end sequencing) is proportionately assigned to the
transcripts it is compatible with, according to the relative transcript
abundances. Then, in the maximization step, relative abundances are
recalculated using the (proportionately) assigned fragment counts. In
other words, the maximization step is the application of Equation
\ref{eq:rpkm} where counts are replaced by expected counts. A key
property of the EM algorithm is that the likelihood increases at every
step \cite{ASCB2005}.
The illustration in Figure \ref{fig:em} shows 2 iterations of the
algorithm and one can observe the convergence of $\rho$ with
$\rho_{\mbox{\tt blue}} = 0.33, 0.27, 0.23, \ldots$. EM theory states that $\rho_{\mbox{\tt blue}}$ will converge to
the value in (\ref{eq:exmle}).
\end{example}
 A derivation of the EM algorithm in the case of the model just described (from the general EM algorithm) is detailed
in \cite{Xing2006}.
\begin{remark}[Rescue for multi-reads]
The ``rescue'' method for multi-reads \cite{Mortazavi2008}
is equivalent to one step of the EM algorithm (this important
observation was made in \cite{B.Li2010}). 
\end{remark}

In some cases least squares has been used instead of the EM algorithm, e.g. in \cite{Bohnert2010,Feng2010,Howard2010}.  Least squares
estimates are equivalent to maximum likelihood estimates under the
assumption that the counts are normally distributed (with equal
variances in \cite{Bohnert2010}, Poisson derived variances in \cite{Feng2010} and possibly with a more complicated
variance-covariance structure in \cite{Howard2010} requiring an
additional estimation procedure). Multivariate normal models
approximate multinomial models well when counts are high but are not
suitable when counts are low. There can be a computational problem
with least squares inference. For example, the problem formulation in
\cite{Howard2010} results in an algorithm requiring matrix inversions
of sizes exponential in the number of transcripts. Furthermore, the
least squares approach requires constrained optimization (quadratic
programming) because relative transcript abundances must
be non-negative. This is done in \cite{Bohnert2010,Feng2010} but not
in \cite{Howard2010} where instead a heuristic is used and negative
estimated values are truncated to zero. 

The method in \cite{Montgomery2010} is similar to the least squares
approaches described above, except that $L_1$ minimization is
performed. Unfortunately, the details of the method are not
sufficiently well explained to permit a direct comparison to other
published methods and therefore it has been omitted from Figure
\ref{fig:hierarchy}. 

The inference approaches of \cite{Jiang2009,Katz2010,Trapnell2010} differ from other
published methods in that they are Bayesian rather than
frequentist. In \cite{Katz2010}, instead of using maximum likelihood to infer $\hat{\rho}$, the
posterior distribution is computed assuming a Dirichlet prior. The
approach of \cite{Jiang2009} and \cite{Trapnell2010} is to use a
Gaussian prior which is chosen according to the MLE estimated using
the EM algorithm (with a variance-covariance matrix based on the
inverse Fisher information matrix from asymptotic MLE theory). An
importance sampling procedure is then used to sample from the
posterior distribution \cite{Liu2008}.

\section{Differential analysis}

Accurate relative transcript abundance estimates are important not just for
comparing isoform abundances to each other, but also for many other
RNA-Seq applications. A common question for which RNA-Seq is now used is
to determine differentially transcribed RNAs (the term
differential expression is frequently used but it is a misnomer and the term {\em
  differential analysis} is more accurate). Clearly accurate
relative transcript abundance estimates are desirable in order to have power to detect
differentially transcribed RNAs.

First we note that the multinomial models and Poisson models are equivalent in terms of
differential analysis by virtue of the following elementary lemma:
\begin{lemma}
\label{lem:Poissoncond}
Suppose $X_1,\ldots,X_n$ are Poisson distributed with rates
$\lambda_1,\ldots,\lambda_n$ respectively. Then the distribution of
$X_1,\ldots,X_n$ conditioned on the sum $\sum_{i=1}^nX_i=N$ is
multinomial.
\end{lemma}
These models, however have been found to be unsatisfactory for
differential analysis because observed counts in technical, and
even more so, biological replicates, do not behave multinomially. This is a common phenomenon observed in count-based
experiments, and is referred to as over- or under- dispersion. Papers
such as \cite{Anders2010,Robinson2010,Srivastava2010,L.Wang2010} have proposed numerous alternatives to the
multinomial model, for example assuming instead that counts are
distributed according to a negative binomial, or generalized Poisson distribution. The differences between the methods
\cite{Anders2010,Robinson2010,L.Wang2010} are mainly in how they
estimate parameters, and a thorough discussion of how they differ is
beyond the scope of this review. A key point, however, is that these approaches
all focus on single-end uniquely mappable reads (i.e., the models in
Section 3) and this has 
the major drawback of not allowing for differential 
analysis of individual transcripts in multi isoform genes, and of
biasing results in gene families. Recent versions of the Cufflinks
software \cite{Trapnell2010} address this problem. 

\begin{remark}[Models, quantification and differential analysis] We
  can now summarize the relevance of model selection for
  quantification and differential analysis as follows: Multinomial
  models and Poisson models are equivalent for quantification and for
  differential analysis (by virtue of Lemma
  \ref{lem:Poissoncond}). Negative binomial models are equivalent to
  multinomial models for quantification, but result in different (more conservative)
  results in differential analysis. Negative binomial models will
  produce different relative abundance estimates if generalized to the
  multi-read case but this has not yet been done. Normal linear models
  in the multi-read case result in different quantification and
  differential analysis results although they can be viewed as an 
  approximation of the Poisson (or multinomial) models.
\end{remark}

Finally, we note that optimized differential analysis of RNA-Seq experiments involves not
only the selection of appropriate models and robust parameters
estimation techniques but also appropriate experimental design \cite{Auer2010}.

\section{Discussion and future directions}

Models of RNA-Seq have advanced greatly in complexity and accuracy in
the three years since the protocol was developed; in this review alone we
have discussed more than 30 different methods that have been
published. At the same time, there has been considerable progress in
RNA-Seq technology, both in protocol development to reduce bias \cite{Levin2010}, and
in sequencing technology that has resulted in much higher throughput
\cite{Mardis2011}. These developments have led to remarkable progress
in the accuracy of RNA-Seq based relative transcript abundance
estimates in the short time since introduction of the technology. Eventually, as
single molecule based technologies mature \cite{Raz2011} and
reads as long as entire transcripts can be produced
\cite{Eid2009}, RNA-Seq will be ``solved'' in the sense that a single
sequencing experiment will directly reveal the transcriptome being queried.

However with present technologies, and for the foreseeable future,
improvements in RNA-Seq modeling are required to best utilize the data
produced in RNA sequencing experiments. Furthermore, related
technologies such as ribosomal profiling \cite{Ingolia2009} require
similar models because they rely on probabilistically assigning short
reads to transcripts, yet they are limited by experimental rather than
technological issues and the types of models we have described are
therefore likely to be relevant for a long time.

A key question, in light of the many remarks we have made, is what
models are relevant in practice, and how a model should be selected in
conjunction with available sequencing technology. We begin with what
we believe is the most important consideration:
\begin{remark}[Read length and model] 
The multi-read models presented in Section 3 that are suitable for
reads which may map to multiple transcripts are essential for accurate
quantification and differential analysis, even in organisms without
splicing or with few multi-isoform genes. This is because of
multi-gene families, and repeated domains, that can result in
ambiguously mapped reads. This issue is particularly pronounced for
short reads ($<50$'bp) \cite[Table 1]{Turro2011}. Longer reads and fragments
mitigate multi-mapping issues due to gene duplicates, however in that case effective length
corrections (Section 2) become very important. Use of Equation
\ref{eq:rpkm} with $l_t$ in the denominator instead of $\tilde{l}_t$
can affect relative abundance estimates by up to 30\% for fragments
with an average size of 200 (as is common in current
protocols). It is important to note that the improvement in mapping of
longer reads may come at the expense of the number of reads, and in
\cite{B.Li2010} it is shown that more short reads are better for
accurate quantification than fewer long reads. However, it should
also be noted that for other RNA-Seq analysis problems,
e.g. transcriptome assembly, long reads and fragments are crucial \cite{Trapnell2010}.
\end{remark}

The simplest models discussed
in this review assume uniformity of fragment location across
transcripts, yet observed data does not conform to this assumption (see, e.g. \cite[Figure 6]{Evans2010}). It is
therefore highly desirable to model bias using more general models, and results in recent papers
such as \cite{Roberts2011} show better agreement between RNA-Seq and
qRT-PCR based estimates when bias is taken into account. Even newer
protocols result in biases \cite{Levin2010} that can be mitigated via appropriate
modeling. However even with current state of the art corrections for
sequence and positional bias, unexplained biases in the data continue
to be observed \cite{J.Li2010,Roberts2011}.  A possible source of bias
that remains to be explored and may affect RNA-Seq experiments is
secondary structure in RNA, and the effect it may have on the
various fragmentation and reverse transcription steps of existing
protocols. Recent progress on sequencing based assays for measuring
structural features of RNA molecules may help to establish a
connection between structure and bias, and to quantify the effect if
it exists \cite{Aviran2011,Kertesz2010,Lucks2011,Underwood2010}.

A crucial area where progress is needed is in the assessment of accuracy of
RNA-Seq. The exact accuracy of relative abundance estimates based on
the techniques reviewed in this paper is currently unknown, and
benchmarks with respect to qRT-PCR \cite{Bullard2010} or nanostring
\cite{Roberts2011} have cast doubt on whether those technologies are more accurate
than RNA-Seq and suitable as ``gold standards''.  Most
importantly, there is a need for systematic
benchmarks where abundances are known {\it a priori}, yet where experiments mimic the
complexities of {\it in vivo} transcriptomes. 

Another aspect of RNA-Seq that has yet to be fully explored is the
connection between relative abundance estimation and transcriptome
assembly. In \cite{Trapnell2010}, it is shown that an incomplete
transcriptome can bias relative abundance estimates. For this reason,
it was proposed that analyses should be performed with respect to an
assembly generated from the data, rather than using a ``reference''
annotation. Indeed, in almost every RNA-Seq study performed to date
many novel transcripts have been detected, even in extensively
annotated species such as Drosophila
\cite{Graveley2011}. Transcriptome assembly is therefore crucial at
the present time for accurate relative
abundance estimation. 

Accurate and complete transcriptome assembly is, in turn, dependent on
the ability to accurately quantify relative transcript
abundances. This is because fragment lengths are currently much shorter than
transcript lengths, and therefore local estimates of relative
abundance are the only information available for phasing
distant exons during assembly  \cite{Wang2008}. For this reason, {\em
  statistical assembly} approaches such as \cite{Laserson2011} need to be
developed for transcriptome assembly. Despite initial work by a number
of groups (personal communication), the problem of how to perform statistical
assembly efficiently and accurately remains open.
\newpage
\section{Acknowledgments}

I thank Adam Roberts and Cole Trapnell for many helpful discussions that clarified our
understanding of RNA-Seq and that led to many of the remarks in this paper. Meromit Singer, during the
course of many discussions with me, questioned the relevance and
interpretation of alternative likelihood formulations for methyl-Seq models;
those conversations led to similar questions about RNA-Seq models, and
finally to the comments on the equivalence between multinomial and Poisson models for
RNA-Seq discussed in Section 4 and Appendix I. Colin Dewey provided
many helpful suggestions and comments after reviewing a preliminary
version of the manuscript. Finally, thanks to
Sharon Aviran, Nicolas Bray, Ingileif Hallgr\'{i}msd\'{o}ttir, Valerie
Hower, Aaron Kleinman, Megan Owen, Harold
Pimentel, Atif Rahman, Adam Roberts, Meromit Singer and Cole Trapnell for valuable
comments and insights during the writing of this paper.

\bibliographystyle{amsplain}
\bibliography{cufflinks}
\newpage
\section*{Appendix I: Equivalence between the paired-end Poisson and
  multinomial models for RNA-Seq}

In this section we show that the general model described in Section 5
is equivalent (in the sense discussed in Section 4) to a Poisson model
for paired-end reads (with the same bias model).

To describe the Poisson model we need some extra notation:
a weak composition of $n$ into $k$ parts is an ordered tuple of
non-negative integers ${\bf c}=(c_1,\ldots,c_k)$ such that
$c_1+\cdots+c_k=n$. If ${\bf c}$ is a weak composition of $n$ we write
${\bf c} \vdash n$. We note that the number of weak compositions of $n$ into $k$ parts is 
\begin{equation}
{n+k-1 \choose k-1} \quad =\quad  {n+k-1 \choose n}
\end{equation}
For convenience, we use the notation $|{\bf c}|$ for the number of
parts in a composition.

We will make use of multinomial coefficients, and we use the
convention that 
\begin{equation}
{n \choose {\bf c}} \quad =\quad  {n \choose c_1,\ldots,c_k} \quad
=\quad  \frac{n!}{c_1!c_2!\cdots c_k!}
\end{equation}
where ${\bf c}\vdash n=(c_1,\ldots,c_k)$ is a composition of $n$ into
$k$ parts.

Finally, we note that a function $P:F_s \rightarrow s$ induces a weak
composition of $|F_s|$ where the composition is given by the
cardinalities of the sets $\{f:P(f)=(t,i)\}$ as $(t,i)$ ranges over
  the elements of $s$. We use $|P|$ to denote this composition.
\begin{example}
Suppose that $S=\{(t_1,4),(t_2,6),(t_3,3)\}$ is an equivalence class
with 3 elements corresponding to three distinct positions in three
different transcripts ($t_1,t_2,t_3$). Suppose that the set of
fragments whose $3'$ ends align to $S$ consists of four fragments:
$F_s=\{f_1,f_2,f_3,f_4\}$. There are $3^4=81$ different functions
$P:F_s \rightarrow S$. If $P(f_1)=P(f_3)=(t_1,4)$ and
$P(f_2)= P(f_4) = (t_3,3)$ then $P$ induces the weak composition ${\bf c} \vdash
4$ with $3$ parts given by ${\bf c} = (2,0,2)$. We denote this weak
composition ${\bf c}$ by $|P|$.
\end{example}

We conclude by highlighting an elementary, yet crucial step in the
derivation of the likelihood function. Suppose that $\{a_{ij}\}_{i=1,j=1}^{n,m}$ are $n \cdot
m$ indeterminates. Then the following two polynomials are equal:
\begin{equation}
\label{eq:keyidea}
\sum_{f:[n] \rightarrow [m]}\prod_{i=1}^n a_{i,f(i)} \quad =\quad 
\prod_{i=1}^n\sum_{j=1}^m a_{ij}.
\end{equation}
The expression on the left in (\ref{eq:keyidea}) consists of $m^n$
monomials, each with $n$ terms, but the factored expression on the right shows
that the polynomial can be evaluated using only $O(nm)$ operations.

Next we turn to the likelihood function.
Let $s \in U = \mathcal{T}/\mathord\sim$ and let ${\bf c} \vdash |F_s|$ be a weak
composition of $|F_s|$. 
We define $\eta_{s,{\bf c}}$ to be the sum
\begin{equation}
\label{eq:alpha}
\eta_{s,{\bf c}} \quad  = \quad \sum_{P:F_s \rightarrow s:|P|={\bf c}} \prod_{f
  \in F_s} \zeta^f_{P(f)} e_{P(f),f}.
\end{equation}
The number $\eta_{s,{\bf c}}$ is the probability of observing the assignment of fragments in $F_s$ to elements of $s$ where the numbers of fragments originating from
each $(t,i) \in s$ is given by $c_{(t,i)}$. 

Our model for RNA-Seq is {\em Poisson}, which means that we assume
that the number of fragments ending at a given site is Poisson
distributed. Specifically, we assume that the number of fragments with
$3'$ end $(t,i)$ is 
\begin{equation}
\lambda_{(t,i)}\quad =\quad \kappa_t \frac{w_{\frac{i}{l_t}}\cdot u_{(t,i)} \cdot  \sum_{j=1}^{i-1}\frac{D(i-j)}{\sum_{k=1}^{i-1}D(i-k)} v_{(t,j)}}{\tilde{l}_t},
\end{equation}
where the notation is the same as that in Section 5. As in Section 4,
we let $\beta_t = \frac{\kappa_t}{N}$.
The likelihood
function we now derive directly generalizes that of \cite{Jiang2009}
to paired-end reads. The likelihood function is
now given by
\begin{eqnarray}
\mathcal{L} & = & \prod_{s \in U} \left(
 \sum_{{\bf c}\,\vdash |F_s|:|{\bf c}|=|s|}  \frac{1}{{|F_s| \choose {\bf
       c}}}  \left(
\prod_{(t,i) \in s} 
\eta_{s,{\bf
    c}}e^{-\lambda_{(t,i)}}\frac{\lambda_{(t,i)}^{c_{(t,i)}}}{c_{(t,i)}!}
\right) \right) \label{eq:lik2} \\
& = & \prod_{s \in U} \frac{1}{|F_s|!}\left(
 \sum_{{\bf c}\,\vdash |F_s|:|{\bf c}|=|s|}  \left(
\prod_{(t,i) \in s} c_{(t,i)}!
\eta_{s,{\bf
    c}}e^{-\lambda_{(t,i)}}\frac{\lambda_{(t,i)}^{c_{(t,i)}}}{c_{(t,i)}!}
\right) \right) \\
& \propto & \prod_{s \in U} \left( \prod_{(t,i) \in s} e^{-\lambda_{(t,i)}}
\right) \left(  \sum_{{\bf c}\,\vdash |F_s|:|{\bf c}|=|s|}  \prod_{(t,i) \in s} \eta_{s,{\bf c}}
  \lambda_{(t,i)}^{c_{(t,i)}} \right)\\
& = & \prod_{s \in U} e^{-\sum_{(t,i) \in s}\lambda_{(t,i)}}
\left(  \sum_{{\bf c}\,\vdash |F_s|:|{\bf c}|=|s|}  \left(
   \sum_{P:F_s \rightarrow s:|P|={\bf c}}
  \prod_{f \in F_s}  \zeta^f_{P(f)}e_{P(f),f} \lambda_{P(f)}\right) \right)\\
& = & \prod_{s \in U} e^{-\sum_{(t,i) \in s}\lambda_{(t,i)}}
\left(  \sum_{P:F_s \rightarrow s}
  \prod_{f \in F_s}  \zeta^f_{P(f)} \lambda_{P(f)}\right) \\
& = & \prod_{s \in U}  e^{-\sum_{(t,i) \in s}\lambda_{(t,i)}}
\left( \prod_{f \in F_s} \sum_{(t,i) \in
    s}\zeta^f_{(t,i)}\lambda_{(t,i)} \right)\\
& = &  \left( \prod_{s \in U} e^{-N\sum_{(t,i) \in s}   \beta_t
    \frac{w_{\frac{i}{l_t}}\cdot u_{(t,i)} \cdot
      \sum_{j=1}^{i-1}\frac{D(i-j)}{\sum_{k=1}^{i-1}D(i-k)}
      v_{(t,j)}}{\tilde{l}_t} }\right) \\
& & \times \left(  \prod_{s \in U} \prod_{f \in F_s} \sum_{(t,i) \in
    s}  \beta_t e_{t,f} \frac{ u_{(t,i)} \cdot v_{(t,i-l+1)} \cdot w_{\frac{i}{l_t}}
  \frac{D(l)}{ \sum_{k=1}^{i-1}D(i-k) } }{\tilde{l}_t}
\right)\\
& =  & e^{-N\sum_{t \in T}   \beta_t}
 \times \left(  \prod_{s \in U} \prod_{f \in F_s} \sum_{(t,i) \in
    s}  \beta_t e_{t,f}\frac{ u_{(t,i)} \cdot v_{(t,i-l+1)} \cdot w_{\frac{i}{l_t}}
  \frac{D(l)}{ \sum_{k=1}^{i-1}D(i-k) } }{\tilde{l}_t}
\right). \label{eq:lik}
\end{eqnarray}
We note that the term in parentheses in (\ref{eq:lik}) is exactly the same expression as
(\ref{eq:genlik}) with $\alpha_t$ replaced by $\beta_t$. As in the derivation in Section 3, it is easy to
see that the likelihood functions are maximized at
$\hat{\alpha}=\hat{\beta}$. 
\section*{Appendix II: Notation}

The notation table below shows the names of all the variables we
use. It is divided into parts using four divisions: the first is
notation for structures independent of an experiment
(transcripts). The second is notation for data from the experiment
(fragments and their alignments). The third is notation for the
parameters of the model. The fourth consists of helper variables that
are calculated from the primary variables in the first three categories.
\begin{table}[!h]
\begin{tabular}{|c|l|}
\hline
$T$ & set of transcripts\\
$\mathcal{T}$ & set of transcripts together with their positions\\
$l_t$ & length of transcript $t$\\
\hline
$F$ & set of fragments\\
$F_s$ & The set of fragments aligning to an equivalence class $s$ of
alignment positions\\
$l_t(f)$ & the length of a fragment alignment to a transcript $t$\\
$X_t$ & number of fragments aligning to a transcript $t$\\
$X_s$ & number of fragments ending (or starting) at a site in $s$\\
{\bf Y} & compatibility matrix (between fragments and transcripts)\\
{\bf C} & compatibility matrix (between sites and transcripts)\\
\hline
$\rho_t$ & abundance of transcript $t$\\
$\lambda_{(t,i)}$ & Poisson rate parameter for position $i$ in transcript $t$\\
$\alpha_t$ & probability of selecting a fragment from transcript
$t$ in the multinomial model\\
$\kappa_t$ & Poisson rate parameter for transcript $t \in T$\\
$\beta_t$ & The probability of selecting a fragment from transcript
$t$ in the Poisson model\\
$e_{t,f}$ & probability of observing $f$ given the sequence in $t$\\
$u_{(t,i)}$ & $3'$ bias weight\\
$v_{(t,i)}$ & $5'$ bias weight\\
$w_{x}$ & positional weight where $x \in [0,1]$\\
$D$ & Fragment length distribution\\
$\tilde{l}_t$ & effective length of transcript $t$\\
\hline
$\eta_{S,{\bf c}}$ & Probability that fragments were sequenced from
position $i$ of transcript $t$\\
$\zeta^f_{(t,i)}$ & Probability of selecting a fragment of length
$l_t(f)$ given $3'$ end $(t,i)$\\
\hline
\end{tabular}
\caption{Notation.}
\end{table}

\end{document}